# Online self-disclosure and wellbeing of adolescents: A systematic literature review

**Mubarak, S.,**
School of Information Technology and Mathematical Sciences,
University of South Australia.
Mawson lakes Campus
Email: sameera.mubarak@unisa.edu.au

**Mubarak, A.R.,**
Department of Social Work,
Flinders University,
Adelaide SA 5001
Email: mubarak@flinders.edu.au

## Abstract

The present research investigated the extent of adolescents' online self-disclosure (SD) through a systematic decadal literature review. This review identified three major research foci, with studies categorized mainly as: (1) factors contributing to online SD of adolescents; (2) risks and consequences of online SD; and (3) future directions of research and practical interventions to address the problems. A detailed examination of the variables covered by the studies indicated that only a few aspects related to adolescents' online SD were addressed but were not analysed in depth. Some aspects of online SD received much attention while others received none. Based on these findings, the present research argues that further research using sophisticated scientific research designs is needed to understand adolescents' online SD.

## 1 Introduction

The concept of self-disclosure (SD) has its origins in the fields of sociology and psychology. It has many psycho-social implications. Disclosing one's identity and personal characteristics to other human beings in order to build strong human relations is a natural human instinct; an urge to connect with others based on mutual disclosure and mutual interests. Human beings may go to the extreme of disclosing what is supposed to be their most sensitive information in order to 'win the heart' of others. While SD has been a natural way of building rapport between human beings since their origin, in recent years the term 'SD' has come under special scrutiny due to the advent of the internet and its radical changes to methods of human interaction. Prior to the internet, SD occurred only between human beings who knew each other and usually within their own community. The internet's capability of easily bringing strangers together to interact has radically changed this scenario. Specifically, internet platforms such as social networking sites (SNS) have created the opportunity of bringing strangers of all age groups together as one social group, at times forcing them to interact with one another as they would with people who live in separate social environments. While adults have the maturity to manage this interactional process, studies are still trying to understand the extent to which adolescents use online SD appropriately on SNS while managing to protect themselves by not over-disclosing and leaving themselves vulnerable to legal and child protection problems. This scenario has brought significant changes to the concept of SD and led to recent research into the concept of online SD from social, psychological, technological and legal points of view.

Adolescents in particular have come under increased scrutiny due to their crucial stage of human development; the transition from childhood to adulthood. Adolescence is well known as an age of transformation; of stress and strain in the process of creating a self-identity. Buhrmester and Prager (1995) argue that SD is a developmental need of adolescents, who use it to gain social input that addresses the underlying issues and concerns typical of their age. Establishing rapport with others is highly important in this process. Many recent research studies have focused on the issue of adolescents' online SD on SNS, but to the researchers' knowledge a systematic review of the literature in this area is yet to be undertaken. Therefore, this study reports a systematic decadal review of the literature about this issue. The review's focus is the exploration of the key concepts reported in the literature related to adolescents' online SD, the consequences of problematic SD, and studies reporting professional interventions for adolescents with SD issues and their outcomes, and addressing gaps in the literature to guide future research and development.

## 2 Context

### 2.1 Online Self-Disclosure (SD)

Online platforms allow users to write and post anything they want, without restrictions on disclosure of personal or sensitive information, such as photos, addresses and other identifying details. Once posted online, this information can be accessed instantly by people who are not meant to receive it. The practice of revealing personal information, referred to as self–disclosure (SD), has been interpreted in many ways. In general, SD refers to the process of making the self known to others (Jourard and Lasakow 1958, p 91). According to Derlega and Grzelak (1979, p 152), SD means any information exchange that refers to the self, including the personal states, dispositions, past events and plans for the future. However, the concept of SD needs to be understood uniquely depending on the background of the person involved. Adolescence is a unique age-group in this regard. It is timely that attempts are made to understand the ways in which adolescents disclose themselves online.

The internet has redefined the concept of SD due to the unique opportunity it presents for users to display a wide range of information about themselves that can be visible to a vast audience depending on a user's privacy settings (Krasnova and Veltri 2011). Hence, SD in the online environment needs to be viewed differently. The online environment enables sharing of someone's sensitive and private personal information without direct communication with and often without the knowledge of the person whose SD is involved. This results in people losing control over their SD and in some situations causing them harm. Specifically, the decision about whom to accept as a friend on SNS and what kind of SD is appropriate depends on users' ability to control their personal information online (Ellison et al. 2011). While Derlega and Grezlak (1979) argue that SD enables identity development through social validation, gaining social control and developing relationships, this process takes on a unique meaning when it occurs in cyberspace, and when children and adolescents are involved.

### 2.2. Online SD and Adolescents

As discussed above, online SD poses several challenges to online users and the community as a whole. Boyd and Ellison (2007) argue that perceived privacy threats are changing the way users disclose information about themselves. Spack-Jones (2003, p 246) observes that information posted online remains permanent in one form or another, which causes the highest concern. Mesch and Beker (2010) argue that the online environment has a generative effect, leading to the formulation of norms that are different from the offline environment. Yet another problem is the disinhibited nature of SD in cyberspace, which can result in cyberbullying and other forms of online harassment for children and adolescents (Valkenburg and Peter 2009). The consequences of online SD impact adolescents to a greater extent than adults because adolescents show a higher tendency to engage in risk-taking behaviours and seem more prone to SD on SNS (Albert and Steinberg 2011). Moreover, adolescents' decision making processes are more stimulated by short-term rewards than long-term perspectives, with adolescents less inclined to evaluate the potential risks of SD (Albert and Steinberg 2011; Taddicken and Jers 2011). The presence of peers on SNS augments the allure of immediate rewards and a reduced focus on the potential costs (Albert and Steinberg 2011). Moreover, by sharing personal information online, adolescents might wish to achieve a more positive and pronounced self-presentation instantly, even if this means violating the traditional social norms of privacy (Jordán-Conde, Mennecke, and Townsend 2014).

## 3 Objectives

The main objectives of this decadal review were to:

    1. Review the pattern of adolescents' online SD, the format of SD, intended audiences of

       SD and motivation behind SD.

2. Evaluate gender, age and personality factors represented in the reviewed articles.

3. Review the risks and consequences of online SD discussed in the articles.

4. Study professional interventions and future research directions to overcome the negative consequences of adolescents' online SD.

# 4  Methods

The systematic review method summarizes information from several studies in order to answer the research question through critical evaluation (Silva et. al. 2012) of the findings, thus bringing together the available research data from independent studies. According to the Cochrane Collaboration, an international organization dedicated to promoting and managing systematic reviews, literature reviews provide more precise estimates of the effects of an intervention than individual studies alone. They allow for investigation of consistencies and differences across studies (Higgins and Green, 2011). Systematic literature reviews also help make sense of larger bodies of information that are available in different forms, and assist in finding answers to questions about what works and what does not. They also map out areas of uncertainty and identify where little or no relevant research has been done (Petticrew and Roberts 2006, p 2).

## 4.1  Research Strategy

Several strategies were used to perform an exhaustive search for literature fitting the inclusion and exclusion criteria detailed below. Additionally, the reference lists of all included articles were scanned for potentially relevant articles. The search for articles was conducted using the electronic databases Science direct, Ebscohost, Engineering Village, IEEE, Proquest, PsychINFO, PschLIT, MEDLINE, EMBASE, ASSIA, Social work abstract, Social science abstract and Google Scholar. The key words used during the search were 'Adolescents', 'Children', 'Youth', 'Online', 'Self-disclosure' and 'Revealing Privacy'. The first search found 2786 relevant articles. After abstract reviewing and evaluating articles according to the inclusion and exclusion criteria below, 29 full text articles were selected.

### 4.1.1  Inclusion Criteria

- The articles' primary focus was to be on adolescents of the age-group 11–19 years and their online SD.
- Articles must be peer reviewed and published in a conference proceeding or a professional journal between January 2004 and December 2014.
- Research methodologies of selected articles were to be qualitative, quantitative and mixed approaches.

### 4.1.2  Exclusion Criteria

- Newspaper articles or book reviews.
- Articles including an adult sample for comparison with an adolescent sample.

## 4.2  Data extraction and validity

Initially, a broad electronic search was conducted using key words. Studies fulfilling the above inclusion criteria were short listed for a review of their abstracts. In the next stage, short listed abstracts were further assessed by quick scanning of the whole content of the article. Once the basic review was completed, repetitions of articles were identified and discarded. In this process, the researchers noted the details of articles which were included and excluded. In each stage, the exact number of articles considered for the present study was noted. All articles using qualitative research methods were thoroughly reviewed using the Critical Appraisal Skills Programme (CASP) (CASP, 2006). Similarly, articles using quantitative research methods were analysed using the Downs and Black instrument (Downs and Black 1998).

# 5  Results

Analysis of the 3 qualitative, 19 quantitative and 6 mixed methods articles found five major themes: motivation; gender and personality; dangers and consequences; interventions; and future directions for research. These are summarised in Table 1.

| Qualitative | Quantitative (Survey) | Mixed/secondary/other |
|---|---|---|
| Williams and Merten (2008) | Chiou (2006) | Mesch and Beker (2010) |
| Davis (2012) | Steijn (2014) | Jordán-Conde et al., (2014) |
| Fox et al., (2010) | Valkenburg and Peter (2005) | Valkenburg and Peter (2011) |
| | Farber et al., (2012) | Hinduja and Patchin (2008) |
| | Blau (2014) | Bryce and Fraser (2014) |
| | Walrave and Heirman (2012) | Bryce and Klang (2009) |
| | Liu et al., (2013) | |
| | Schouten et al. (2007) | |
| | Peter et al., (2005) | |
| | Cho (2007) | |
| | Valkenburg et al., (2011) | |
| | Christofides et al., (2012) | |
| | Ong E.Y.L et al., (2010) | |
| | Ji et al., (2014) | |
| | Heirman et al., (2013) | |
| | Courtois et al., (2012) | |
| | Liu and Brown (2014) | |
| | Valkenburg and Peter (2009) | |
| | Liu (2014) | |
| | Van Gool et al., (2015) | |

*Table 1. Types of articles included in the present study*

## 5.1 Motivation behind adolescents' online SD

Social development and anonymity were two of the most common topics in the literature reviewed. Valkenburg et al. (2011) observed that the concept of social development played a crucial role in adolescents' SD in cyberspace. These authors argue that the online environment provides adolescents with an experimental space in which to try out new things as a part of their social development. They also suggest that SD may be beneficial to adolescents by enhancing closeness with peers (Valkenburg and Peter 2011). Similarly, Conde and Mennecke (2014) report a direct relationship between psychosocial development and adolescents' disclosure of intimacy in online platforms, while Ji et al. (2014) and Jordan-Conde et al. (2014) argue that online privacy disclosure with peers helps adolescents establish more friendships, which enables their psychosocial development. Liu and Brown (2014) demonstrate that adolescents' SD on SNS has a direct association with social capital. Walrave and Heirman (2012) argue, however, that the urge to gain a sense of belonging and to become popular on SNS pushes adolescents to self-disclose more than necessary. These authors also argue that existing offline contacts and trust with other users make adolescents reveal more about themselves. In this study, it was observed that social anxiety decreased adolescents' SD. Wang et al. (2014) report that less emotionally secure respondents engage less in online SD.

Anonymity, which has been made so easy in cyberspace, was the focus of many of the articles included in this review. Chiou (2006) argues that the faceless atmosphere of cyberspace encourages greater sexual SD by adolescents online. Walrave and Heirman (2012) argue that the anonymity offered by the online environment may be one of the factors that encourages adolescents to self-disclose beyond safe limits. Christofides and Muise et al. (2012) report that online SD is unrelated to the importance of information control. Instead, they argue that the anonymous online platform may open adolescents to certain negative experiences that occur during online interactions, but they keep these to themselves and do not share them with parents or other authority figures. In some studies, such as Fox et al. (2010), online SD has been used to enhance the outcomes of medical treatments for things such as chronic skin conditions. Thus, the studies included in the review represent both positive and negative outcomes of adolescent online SD. However, none of the literature involved deeper psycho-social analysis of adolescent online SD and its relevance to social development.

### 5.2 Gender and personality factors associated with SD

A detailed review of the above articles revealed gender and personality factors repeatedly attracted research attention when studying online SD. While discussing gender participation in SD behaviour, Chiou (2006) notes that more males than females engage in SD. This finding is supported by Valkenburg et al. (2011), whose study reveals that more boys engage in online SD than girls. Davis (2012), however, reports that girls engage more in online SD than boys. In regard to the depth of SD, it was observed that female adolescents disclose less during SD compared to male adolescents (Ji, Wang, et al. 2014). Contrary to these findings, Mesch and Beker (2010) found no gender difference in SD. Thus, the studies included in the review represent both positive and negative outcomes of adolescent online SD. However, none of the literature involved deeper psycho-social analysis of adolescent online SD and its relevance to social development..

While observing the relationship between personality and SD, most of the studies tried to reinforce their argument that extroverted adolescents engaged in revealing their fashionable and glamorous self. Extrovert and narcissistic personalities seem to have attracted more research attention than other personality types, with studies observing that extrovert and narcissistic personality types often display narcissistic ideas in their Facebook profiles by displaying their physical attractiveness. For example, Ong et al. (2011) observe that narcissistic personalities update their status more frequently. These authors argue that narcissism increases adolescents' online SD. Similarly, Peter et al. (2005) argue that adolescents who are 'high on narcissism' tend to post all kinds of information about themselves. It was observed that extroverted adolescents engaged in online SD and made new friendships more frequently compared to introverted adolescents, resulting in a phenomenon termed 'rich get richer.' At the same time, studies such as that by Schouten (2007), suspect that there may not be a direct correlation between personality characteristics and online SD. The lack of studies exploring the long-term implications of gender and personality on adolescent online SD, and the cross-sectional methodology of the reviewed literature, resulting in mixed findings, provide inconclusive evidence about the effects of gender and personality on adolescents online SD.

| Authors | Gender and personality factors associated with SD |
| --- | --- |
| Chiou (2006) | More males than females engage in self-disclosure |
| Liu et al. (2013) | Narcissism increases adolescents' PII disclosure |
| Peter et al. (2005) | Self-disclosure and new friendship means Rich get richer |
| Davis (2012) | Girls are more engaged in online SD |
| Valkenburg et al. (2011) | More boys engaged in online SD than girls |
| Ong E.Y.L et al. (2010) | Narcissistic personalities updated their states more frequently |
| Ji et al., (2014) | Female adolescents disclose less compared to male adolescents |
| Schouten et al. (2007) | Personality is indirectly related to self-disclosure |

*Table 2. Gender and personality factors associated with SD*

### 5.3 Dangers and Consequences of SD

At least 10 studies included in the present literature review reported that online SD had several dangers or negative consequences for adolescents. Increased anonymity in online spaces creates greater SD by adolescents resulting in them posting intimate and candid personal information (Williams and Merten, 2008). Steijn (2014) reports that blogs posted by adolescents contain sensitive information such as their images and location, making them an easy target for internet predators. Wang et al. (2011) also report that adolescents' frequent display of sensitive information, such as names and photos, makes them easy targets for abuse. Half of the sites analysed for Steijn's study could potentially jeopardize the identity security of the adolescent participants. Blau (2014) observes that adolescents had frequently engaged in risky SD behaviours such as sending photos and posting personal details. Similarly, Walrave and Heirman (2012) report that adolescents apply more lenient privacy settings and are less concerned about the consequences of the risks they take. These authors also argue that increased trust in other users influences adolescents taking risks when using online SD.

Valkenburg and Peter (2011) explain that some of the consequences of adolescents' risky SD behaviours were online sexual solicitation, sexting, cyberbullying, online harassment and internet addiction. Hinduja and Patchin (2008) report that adolescents also provide clear physical descriptions, which school they attend, and pictures of themselves in swimsuits and revealing clothes, which attract cyberbullies and sexual predators. Bryce and Fraser (2014), in observations from their focus group study, found that adolescents routinely post their name, images, details of appearance and interests online. Surprisingly, participants who had experienced bullying and other problems acknowledged the benefits of SD and ignored the dangers. Bryce and Klang (2009) argue that adolescents experience confusion between their private and public self in online spaces. It is also noted that sometimes adolescents' SD takes place in exchange for incentives offered by commercial website writers (Heirman et al. 2013).

Christofides et al. (2012) observe that adolescents are more likely to disclose more personal information than adults during online SD. Their study shows that awareness of the consequences of online SD was same for adults and adolescents, and that increased time with Facebook or other social media results in increased online SD. Surprisingly, the focus group and survey results of Christofides et al.'s study show that the study samples were confident about the safety of information they published and rejected the notion of information security risks. Similarly, the study by Jordán-Conde et al. (2014) supports the view that SNS users do not consider privacy and the huge ramifications of SD as serious matters. Jordán-Conde et al. argue that this trend can be dangerous. Faber et al. (2012) observe that online SD can be extremely abrupt and limits emotional expression, which many users find overwhelming. The lack of consistency shown in the research findings about adolescents' engagement in risky behaviours while interacting socially with others points to the need for longitudinal studies to help explain why adolescents engage in risky behaviours and how to prevent them.

| Authors | Dangers and Consequences of SD |
| --- | --- |
| Steijn (2014) | Adolescents share their private information |
| Williams and Merten (2008) | SD can be an easy target for internet predators |
| Walrave and Heirman (2012) | Seeks more autonomy, hence indulging in risky behaviours |
| Valkenburg and Peter (2011) | Online sexual solicitation, sexting, cyberbullying, online harassment and internet addiction |
| Hinduja and Patchin (2008) | Attract cyberbullies and sexual predators |
| Bryce and Fraser (2014) | Acknowledged the benefits of SD and ignored the dangers |
| Bryce and Klang (2009) | Experience confusion between their private and public self in online spaces |
| Christofides et al. (2012) | Disclose more personal information than adults during online SD |
| Jordán-Conde et al. (2014) | SNS users do not consider privacy and the huge ramifications of SD |
| Faber et al. (2012) | Extremely abrupt and limits emotional expression |

*Table 3. Dangers and Consequences of SD*

## 5.4  Interventions

Only 9 studies mentioned the need for interventions to reduce the frequency of adolescents' online SD and possible areas of intervention, but none made any concrete efforts to intervene.  William and Merten (2008) highlight the importance of helping researchers/educators and parents to understand the dangers of adolescents' online communication through meaningful discussions regarding conflict in the areas of propriety, ethics and safety in online environments. Supporting this point, Walrave and Heirman (2012) say "being critical" towards SD is essential, and suggest  sensitizing campaigns to increase awareness of audience management and SNS providers implementing safer social networking principles, such as privacy features by default. They reiterate the fact that creation of awareness about e-safety issues among parents, teachers and adolescents can create deeper reflection on the disclosure of personal information, which may reduce online risks. Christofides et al. (2012) claim that the negative consequences of online SD are not limited to adolescents and that adults also need to be careful with privacy settings. Thus, adult and adolescent groups should be equally involved in education about safe online SD. Liu et al. (2013) have demonstrated that awareness creation regarding the importance of privacy during online SD significantly reduces risky online SD behaviours among adolescents.

Jordan-Conde et al. (2014) and Valkenburg and Peter (2011) highlight that SNS designers should be listening to the advice of users as these users become more informed about the ways in which technology can serve their social and relational needs while they explore their identities online. These authors also note that enhancing communication and creating awareness among adolescents regarding the risks involved in online SD will reduce the associated risks. Ji et al. (2014) argue that education and awareness creation among adolescents regarding the importance of privacy and security while interacting through SNS are necessary. Hinduja and Patchin (2008) suggest that, as preventive steps, supervising adolescents' online activities and creating awareness about permanency of their online profile in cyberspace are important.  In addition, Bryce and Klang (2009) emphasize the importance of online providers helping their clients to understand the terms of their services and other policy matters prior to using SNS. Similarly, Christofides et al. (2009) suggest that policy makers, practitioners or parents need to stress the possible opportunities and risks involved when disclosing personal information during online SD. The present review of literature identified a clear gap in the understanding of adolescent SD online because most of the studies were cross-sectional in nature and did not attempt to intervene and analyse the impact of intervention on adolescents' online SD. Often, authors made anecdotal comments, pointing their fingers at families or policy makers or schools without any substantial evidence based on in-depth analysis or intervention.

| Authors | Interventions |
| --- | --- |
| William and Merten (2008) | Helping researchers/educators and parents to understand the dangers |
| Walrave and Heirman (2012) | Sensitizing campaigns to increase consciousness |
| Christofides et al. (2012) | Adult and adolescent groups should be equally involved in education |
| Liu et al. (2013) | Awareness creation regarding the importance of privacy |
| Jordan-Conde et al. (2014) | Social network designers should listen to the advice of users |
| Valkenburg and Peter (2011) | Risk reduction by creating Awareness |
| Ji et al. (2014) | Awareness about Privacy and Security |
| Hinduja and Patchin (2008) | Preventive steps and Supervising |
| Bryce and Klang (2009) | Terms of services policy must be strictly enhanced |

*Table 4.  Interventions*

## 5.5  Future directions for further research

This literature review raises several key points in relation to directions for future research. Chiou (2006) claims that future research on pathological SD and resulting interventions would be useful. Valkenburg and Peter (2005) stress the urgency of longitudinal studies on the relationship between online communication and the quality of adolescent interpersonal relationships. Blau (2014) suggests that online users can learn a great deal about the risks involved in online SD if they are provided with an opportunity to explore the risks involved in off-line SD and then compare the consequences of online and off-line SD. Schouten et al. (2007) stress that further research to investigate the long-term consequences of online SD is needed to reveal findings that could provide the basis for interventions.

Cho (2007) stresses the need for more precise SD scales to test Computer Mediated Communication (CMC) characteristics such as anonymity, de-individualization or hyper social effects. Cho also states that direct observation of adolescents' digital media use may provide good scope in terms of knowing more about their online SD behaviour. Ji et al. (2014) claim that to get a real outcome, data mining methods may be useful in collecting real life SNS behaviours for behavioural or semantic analyses. Bryce and Fraser (2014) note that evaluation of risk and trust, and their influence on behaviour in digital environments using qualitative and quantitative methods, would be appropriate. Davis (2012) and Jordán-Conde et al. (2014) emphasize that parents, educators and counsellors need to be provided with interventions to help them understand the importance of assisting adolescents with their online SD. Ross et al. (2009) (article 29) suggest that qualitative research exploring the different motives that influence the perceptions of adolescents' online SD and antecedents that influence their sharing of personal information would be beneficial.

In regard to personality factors, in this review the 5 studies that focused on personality factors associated with online SD all focused on narcissistic and extrovert adolescent personalities. Other personality types did not attract much research attention. In spite of this, Ong et al. (2011) suggest that more research is needed on narcissistic self-representation attributes related to online SD. Valkenburg and Peter (2011) argue that many studies on personality factors related to online SD do not pay adequate attention to sampling and the limitations of these studies in generalizing the results. They also suggest that different types of internet use must be distinguished conceptually. Thus, it is evident that even though researchers realised the urgency of consistency in research and deeper analysis of adolescent online SD as early as 2005, the present literature review suggests that this gap in the research still persists in 2015.

| Authors | Future directions for further research |
| --- | --- |
| Chiou (2006) | Pathological self-disclosure and intervention |
| Williams and Merten (2008) | Investigating specific behaviours in the context of real life |
| Valkenburg and Peter (2005) | Online communication and quality of adolescent relationships |
| Blau (2014) | Exploring offline problematic uses |
| Schouten et al. (2007) | Long term consequences of online self-disclosure |
| Cho (2007) | More precise self-disclosure scales to test anonymity, de-individualization or hyper social effect |
| Davis (2012) | Direct observation of adolescents digital media use. |
| Jordán-Conde et al., (2014) | Parents, educators, counsellors needs to provide intervention. |
| Ong E.Y.L et al., (2010) | More research on narcissistic self-representation based on other self-presentation attributes |
| Valkenburg and Peter (2011) | Different types of internet use must be conceptually distinguished. |
| Ji et al., (2014) | Data mining methods in collecting real life SNS behaviours for behavioural or semantic analyses |
| Bryce and Fraser (2014) | Evaluation of risk and trust in digital environments |

*Table 5. Future directions for further research*

# 6   DISCUSSION

This decadal review consisted of analysis of 29 studies conducted between the years 2004 and 2014 on the topic of adolescents' online SD while using SNS. When the term 'self-disclosure' was used in general, the literature search indicated 2786 studies, highlighting the importance the research community has assigned to the concept of SD. However, when the search was narrowed down to adolescents' online SD, only 29 studies qualified for inclusion. This trend clearly highlights the urgency of more research to understand the processes of adolescents' online interaction and SD. Adolescence as an important milestone in the human development lifecycle has received much research attention in recent years. However, this review clearly indicates limited research studies focusing on adolescent behaviours while using SNS. These online platforms have dominated adolescent life in recent years. Their influence in shaping adolescents' lives is beyond doubt but research attempts to understand adolescents' interaction within SNS is inadequate. This area begs more research attention urgently.

The research studies included in the review cover a range of topics related to adolescents' online SD. The majority of the studies (68%) used quantitative research methods such as surveys, while 21% used a combination of quantitative and qualitative research methods, and only 11% used purely qualitative research methods such as focus group discussions. While each of these research methods has its particular advantages and disadvantages, depending on the area of research, the review highlights that more studies are needed that use qualitative research methods to analyse online SD. The review has revealed that individuals engage in SD to satisfy a range of social needs; SD is a sociological process in which social interaction occurs in a social context. Such a process may be well analysed using in-depth, qualitative, social analytical techniques such as one-to-one personal interviews and focus group discussions. Furthermore, this literature review has highlighted the urgent need for longitudinal studies analysing the psycho-social variables related to adolescents' online SD and how these variables change as teenagers progress from early to mid to later teen years.

When analyzing the background characteristics of adolescents using online SD, the reviewers found that gender and personality characteristics were included in the majority of studies reviewed but there was a level of inconsistency pertaining to the SD of specific gender groups. More research is needed in this area because it appears that the studies include in the present review highlight that SD may be gender specific but this observation remains inconclusive. When designing interventions to enhance the quality of adolescents' online SD, population-specific interventions need to be considered because they may make interventions more successful. Similarly, personality characteristics received wide research attention but this review suggests that only two aspects of personality (extrovert and narcissistic personalities) received overwhelmingly more attention than others. This finding opens up other personality characteristics as possible areas of future research.

The social development of adolescents as a part of SD received considerable research attention and the current evidence suggests that the urge to belong and be popular among peers on SNS pushes adolescents to self-disclose more than is wise. However, the studies reviewed have a narrow focus on only a few variables, such as a sense of belonging and the urge to become popular among peers. Other important variables, such as the influence of family, real life peers and the school environment have received limited research attention. This literature review clearly highlights the urgency of intervention studies focusing on the roles played by the social environment of adolescents in their online SD.

The variables anonymity and negative consequences of SD received consistently wide research attention. The majority of studies reviewed analysed these variables from many different angles. They argue that adolescents engage in many risk-taking behaviours while disclosing themselves in cyberspace. This opens up many other research questions related to other variables that play a role in adolescents' online SD. For example, the term anonymity in cyberspace remains insufficiently defined in the literature reviewed for this research. It is evident that more research is urgently needed to understand the social variables related to cyberspace and what kind of social forces are operating within this world to understand concepts such as self, anonymity, and "others" in cyberspace. It can be argued that without a clear understanding of these fundamentals of cyberspace, it is misleading to conclude that any online behaviour as risky. However, the rapidity with which younger generations have embraced IT tools and their enthusiasm for becoming "netizens" requires researchers to examine in depth the social aspects of the world of cyberspace and how different age-groups behave within it.

One of the aims of this review was to find out whether an evidence base exists for specific types of interventions to help adolescents develop positive online SD behaviours. The review clearly reveals a gap in the research literature in this area. While many studies suggest the urgent need for the creation of empirical evidence to demonstrate the effectiveness of interventions with adolescents, none of them propose any specific research methods for doing so. Further, in spite of research evidence suggesting this gap in the literature since 2004, to the reviewers' knowledge no research study has focused on this area. Research efforts using refined research methods, such as a control group and an experimental group, and blind selection of research samples, are urgently needed so that society can develop a refined understanding of the types of adolescent online SD behaviours and effective interventions to assist adolescents with positive online SD. The literature review regarding future research directions reveals a similar trend. Longitudinal studies to understand adolescent online SD and ways of preventing negative online SD were absent and urgently needed.

## 7  Conclusion

This research investigated the extent of online SD among adolescents through a systematic review of the literature about adolescents' online SD. It has identified three major areas of focus in published research studies that met the inclusion criteria: (1) factors contributing to online SD of adolescents; (2) risks and consequences of online SD; and (3) future directions of research and practical interventions to address the problems. A detailed examination of variables covered by the studies indicates that further research is needed to understand adolescents' use of online SD. Critical analysis of future research directions implies the urgent need for studies using advanced research methods, such as longitudinal research and qualitative studies, exploring social aspects of adolescents' online SD. Policy making was revealed as an important step forward in helping the community assist adolescents in positive SD. Further research is urgently needed to create a research evidence base that will help policy makers understand the ways in which young people can be helped to maximize the benefits of cyberspace while minimizing the risks.